\documentclass[a4paper,12pt]{article}
\pdfoutput=1
\usepackage{jheppub}
\usepackage{hyperref}
\hypersetup{colorlinks=true}
\usepackage[T1]{fontenc} 
\title{Neumann-Rosochatius system for strings in ABJ Model}
\author[a]{Adrita Chakraborty}
\author[b]{Kamal L. Panigrahi}
 \affiliation[a]{Centre For Theoretical Studies, Indian Institute of Technology Kharagpur-721302, India}
\affiliation[b]{Department of Physics, Indian Institute of Technology Kharagpur-721302, India}
\emailAdd{adimanta09@iitkgp.ac.in}
\emailAdd{panigrahi@phy.iitkgp.ac.in}
\abstract{Neumann-Rosochatius system is a well known one dimensional integrable system. We study the rotating and pulsating string in $AdS_4 \times \mathbb{CP}^3$ with a $B_{\rm{NS}}$ holonomy turned on over $\mathbb{CP}^1 \subset \mathbb{CP}^3$, the so called Aharony-Bergman-Jafferis (ABJ) background. We observe that the string equations of motion in both cases are integrable and the Lagrangians  reduce to a form similar to that of a deformed Neumann-Rosochatius system. We find out the scaling relations among various conserved charges and comment on the finite size effect for the dyonic giant magnons on $R_{t}\times \mathbb{CP}^{3}$ with two angular momenta. For the pulsating string we derive the energy as function of oscillation number and angular momenta along $\mathbb{CP}^{3}$.}

\begin{document}
\maketitle
	\flushbottom
\section{Introduction}
Planar integrability of both gauge theory as well as string theory has played a vital role in understanding the celebrated
AdS/CFT duality conjecture\cite{Maldacena:1997re}\cite{Gubser:1998bc}\cite{ Witten:1998qj}in a better way. In this context, 
it was first observed by Minahan and Zarembo\cite{Minahan:2002ve} that the one-loop dilatation operators of the $SU(2)$ sector of $N = 4$ Supersymmetric Yang Mills (SYM) theory can be identified with the Hamiltonian of the Heisenberg spin chain \cite{Beisert:2003jj}\cite{Beisert:2003yb}. As proving the duality for all values of 
coupling is extremely hard, the semiclassical string states in the gravity side have been used to look for suitable gauge theory operators on the boundary. The usual $AdS_5/CFT_4$ duality has been generalized to $AdS_3/CFT_2$ in the presence of mixed NS-NS and R-R flux as well. The sigma model for the string in $AdS_3 \times S^3 \times T^4$ in this mixed flux background has been proved to be classically integrable \cite{Cagnazzo:2012se}. The background solution has further been shown to satisfy the type IIB supergravity field equations, provided the parameters associated to field strengths of NS-NS flux (say $q$) and R-R flux (say $\hat{q}$) are related by the constraint $q^2 + \hat{q}^2 = 1$.
This $AdS_3 \times S^3 \times T^4$ background with 'mixed' three-form fluxes has been an interesting testing laboratory for proving $AdS_3/CFT_2$ duality in the presence of fluxes. This background is conjectured to be originated from the near horizon geometry of the intersecting $(F1-NS5-D1-D5)$ branes in supergravity, although an explicit construction is yet to be found.\\
In adding further examples of the AdS/CFT duality, ABJM theory \cite{Aharony:2008ug} has been conjectured to be dual to the $M$-theory on AdS$_4 \times$ S$^7/ Z_k$ with $N$ units of four-form flux, which for $k << N << k^5$ can be compactified down to a 10 dimensional type IIA string theory on AdS$_4 \times {\mathbb {CP}}^3$, with $k$ being the level of Chern-Simon (CS) theory
with gauge group $U(N) \times U(N)$. The ABJ model \cite{Aharony:2008gk} is an interesting extension of the above with the gauge group $U(M)_{k}\times {\overline{U(N)}}_{-k}$ and the amount of maximal supersymmetry
remaining fixed. The corresponding string dual in the limit of $k\ll N\ll k^{5}$ is conjectured to be type IIA superstring theory on $AdS_{4}\times \mathbb{CP}^{3}$ background with a NS-NS two form holonomy over $\mathbb{CP}^{1}\subset
\mathbb{CP}^{3}$. Using the integrability property of the classical string-sigma model, it is relevant to find generic string solutions and their corresponding field theory duals. It has been proved that the ABJ theory is integrable
both in nonplanar and planar limits \cite{Bak:2008vd, Minahan:2009te,deMelloKoch:2012kv,Mohammed:2012xk} similar to 
its gravity dual. In understanding the string geodesics better, several classes of rigidly rotating and pulsating
string solutions in the background of $AdS_{4}\times\mathbb{CP}^{3}$ with and without flux has been studied, e.g. in 
\cite{Grignani:2008te,Astolfi:2008ji,Ahn:2011nm,Ahn:2008wd,Ahn:2008hj,Jain:2008mt}.

To this end, \cite{Arutyunov:2003za,Arutyunov:2003uj, Kruczenski:2006pk} provided a novel way to find out a large class of simple rotating string solutions by solving a very renowned one dimensional integrable system, known as Neumann model that describes an oscillator on a sphere. The Neumann-Rosochatius (NR) system on the other hand depicts a particle on a sphere with an additional centrifugal potential (proportional to $\frac{1}{r^{2}}$). Reduction of the string-sigma model on $AdS_5 \times S^5$, $AdS_4 \times {\mathbb {CP}}^3$, etc. to the NR system has been quite useful in unravelling new relationships between the integrable structures of the two sides of the AdS/CFT
duality\cite{Hernandez:2017raj,Hernandez:2014eta,Hernandez:2015nba,Ahn:2008hj,Hernandez:2018gcd,Arutyunov:2016ysi}. Such an approach has previously been used to study the finite size effect to the giant magnon (GM) solutions in $AdS_{4}\times S^{7}$ background\cite{Ahn:2008gd}.  It is worth emphasizing that the NR integrable system is very effective in dealing with the classical strings in $R_t \times \mathbb{CP}^{3}$.  Indeed  in  \cite{Ahn:2008gd}, the string dynamics on $AdS_4 \times {\mathbb {CP}}^3$ has been studied by using the NR integrable system. The dispersion relation and subsequently the finite size effects for the giant magnon and single spike solutions for the string on $R_t \times {\mathbb {CP}}^3$ with two angular momenta have been studied in detail. In this article, we wish to extend the analysis presented in \cite{Ahn:2008gd} to the case of spinning closed strings in $AdS_{4}\times \mathbb{CP}^{3}$ in the presence of $B_{NS}$ holonomy. \\

Among various classes of semiclassical strings, the pulsating strings have much better stability than the non-pulsating ones \cite{Khan:2005fc}. The pulsating string concept was first introduced in \cite{Gubser:2002tv}, where it was shown that they correspond to certain highly excited sigma model operators. However unlike rotating strings, pulsating strings are less explored. These solutions were first introduced in \cite{Minahan:2002rc} and further generalized in \cite{Engquist:2003rn}, \cite{Dimov:2004xi}, \cite{Smedback:1998yn}. They have also been explored in $AdS_5 \times S^5$, for e.g. in \cite{Khan:2003sm},\cite{Arutyunov:2003za},  in $AdS_4 \times \mathbb {CP}^3$, for e.g. in \cite{Chen:2008qq}, \cite{Dimov:2009rd}. We wish to show that the pulsating string in $AdS_4 \times \mathbb {CP}^3$ with $B_{NS}$ holonomy can also be reduced to a NR system. We derive the energy of such pulsating strings as a function of the  oscillation number and the angular momenta along $\mathbb{CP}^3$.
    
The rest of the paper is organised as follows. In section 2 we study classical string action in $AdS_{4}\times \mathbb{CP}^{3}$ background in the presence of $B_{NS}$ holonomy and the corresponding NR system. In section 3 we study of rigidly rotating strings in $R_t \times \mathbb{CP}^3$ and compute finite size effect for the dyonic giant magnon solution with two angular momenta. Section 4 is devoted to the study of pulsating string in $R_{t}\times \mathbb{CP}^{3}$. We conclude in section 5 with a brief discussion of our results. 
\section{String in $AdS_{4} \times \mathbb{CP}^3$ with two form NS-NS flux}
We start by writing down the Polyakov action for the bosonic string in the form 
\begin{eqnarray}
S=-\frac{T}{2}\int d\tau d\sigma \sqrt{-\gamma}
\gamma^{\alpha\beta}G_{\alpha\beta}-\frac{T}{2}
\int d\sigma d\tau \epsilon^{\alpha\beta}B_{MN}\partial_\alpha X^M\partial_\beta X^N \ , \nonumber \\
\end{eqnarray}
where $G_{\alpha\beta}=G_{MN}\partial_\alpha X^M\partial_\beta X^N$, $X^N(\tau,\sigma), M, N =0,\dots,9$ are embedding
coordinates of the string. Further, $\partial_\alpha\equiv \frac{\partial}{\partial \sigma^\alpha}\ , \sigma^0=\tau,\sigma^1=\sigma$ and $T$ is the string
tension. The string is embedded into ten dimensional background with the metric $G_{MN}$  and NS-NS two form field $B_{MN}$. Finally
$\gamma_{\alpha\beta}$ is two dimensional world-sheet metric whose equations of motion have the form 
\begin{equation}\label{eqgamma}
T_{\alpha\beta}=-\frac{2}{\sqrt{-\gamma}}\frac{\delta S}{\delta \gamma^{\alpha\beta}}=
-\frac{T}{2}\gamma_{\alpha\beta}\gamma^{\gamma\delta}G_{\gamma\delta}+TG_{\alpha\beta}=0 \ . 
\end{equation}
The supergravity dual background of so called ABJ theory is $AdS_{4}\times \mathbb{CP}^{3}$ background with $B_{NS}$ flux turned on $\mathbb{CP}^{1} \subset \mathbb{CP}^{3}$. The metric is given as
\begin{equation}
ds^{2}=G_{MN}d x^{M}dx^{N}=R^{2}\left(\frac{1}{4}ds^{2}_{AdS_{4}}+ds^{2}_{\mathbb{CP}^{
3}}\right) \ ,
\end{equation}
which in terms of the background coordinates assumes the following form
\begin{eqnarray}
\begin{split}
ds^2 &= \frac{R^{2}}{4}[-\cosh^{2}\rho dt^{2}+d\rho^{2}+\sinh^{2}\rho(d\eta^{2}+\sin^{2}\eta d\chi^{2})] \nonumber \\
&+ R^{2}[d\xi^{2}+\cos^{2}\xi \sin^{2}\xi(d\psi+\frac{1}{2}\cos\theta_{1}d\phi_{1}-\frac{1}{2}\cos\theta_{2}d\phi_{2})^{2}
+ \nonumber \\
+ &\frac{1}{4}\cos^{2}\xi(d\theta_{1}^{2}+\sin^{2}\theta_{1}d\phi_{1}^{2})+\frac{1}{4}\sin^{2}\xi(d\theta_{2}^{2}+\sin^{2}\theta_{2}d\phi_{2}^{2})],
\end{split}
\label{ads4}
 \end{eqnarray}
accompanied by the following NS-NS $B$-field as 
\begin{eqnarray}
B_{NS}= - \frac{b}{2}\Big(\sin2\xi d\xi\wedge (2d\psi+\cos\theta_{1}d\phi_{1}-\cos\theta_{2}d\phi_{2}) +\cos^{2}\xi \sin\theta_{1}d\theta_{1}\wedge d\phi_{1} \nonumber \\
 +\sin^{2}\xi \sin\theta_{2} d\theta_{2}\wedge d\phi_{2}\Big).
 \end{eqnarray}
In addition to the above form of metric and NS-NS flux, there is a dilaton field and Ramond-Ramond two form and four form flux respectively, whose detailed forms are not needed in what follows. When taking $\alpha'= 1 $, the curvature radius $R$ is given by $R^2= 2^{5/2} \pi \lambda^{1/2}$, which is precisely the same as that of ABJM theory.
 
\subsection{Constraints and NR integrable system }
It is convenient to describe string moving in this background
with the help of the embedding coordinates $Y^i$ for $AdS_4$
and $X_i$ for $\mathbb{CP}^3$, where  the embedding  coordinates
describing the background must satisfy the following constraints (see for example \cite{Grignani:2008te}). For $AdS_{4}$ part, the constraint is 
 \begin{equation}
 \sum_{i,j=0}^{4}\eta^{ij}Y_{i}Y_{j}+\frac{R^{2}}{4}=0,
 \end{equation}
 while for $\mathbb{CP}^{3}$, we have the following constraints  
 \begin{equation}
 \sum_{i=0}^{8}{X_{i}}^{2}-R^{2}=0, \quad  \sum_{i=1,3,5,7}(X_
 {i}\partial_{\alpha}X_{i+1}-X_{i+1}\partial_{\alpha}X_{i})=0 \ . 
 \end{equation}
In case of $AdS_4$, the embedding coordinates are related to the global ones by 
\begin{eqnarray}
&& Y_{0}+iY_{4}=\frac{R}{2}\cosh{\rho}e^{it}, \quad Y_{2}+iY_{3}=\frac{R}{2}\sinh{\rho}\sin{\eta}e^{i\chi} \ , \nonumber \\
&& Y_1 = \frac{R}{2}\sinh \rho \cos \eta, \quad  Y_{1}+i\sqrt{(Y_{2}^{2}+Y_{3}^{2}})
= \frac{R}{2}\sinh{\rho}e^{i\eta}  
\end{eqnarray}
while in case of $\mathbb{CP}^3$ we have
\begin{eqnarray}
& &     X_{1}+iX_{2}=\frac{R}{\sqrt{2}}\sin{\xi}e^{i\theta_{1}}, \quad X_{3}+iX_{4}=\frac{R}{\sqrt{2}}\cos{\xi}e^{i\theta_{2}} \ , \nonumber \\
& &    X_{5}+iX_{6}=\frac{R}{\sqrt{2}}\sin{\psi}e^{i\phi_{1}},
\quad 
X_{7}+iX_{8}=\frac{R}{\sqrt{2}}\cos{\psi}e^{i\phi_{2}} \ .  \nonumber \\
\label{constraint2}
\end{eqnarray}For simplicity, we are interested in the string dynamics on $R_{t}\times\mathbb{CP}^{3}$ subspace, which can be
achieved by putting $Y_{1}=Y_{2}=Y_{3}=0$. The metric (\ref{ads4}) in this case, takes the following form 
\begin{equation}
     \begin{split}
         ds^2=&\frac{R^{2}}{4}[- dt^{2}]+R^{2}[d\xi^{2}+\cos^{2}\xi \sin^{2}\xi(d\psi+\frac{1}{2}\cos\theta_{1}d\phi_{1}-\frac{1}{2}\cos\theta_{2}d\phi_{2})^{2}+\\&\frac{1}{4}\cos^{2}\xi(d\theta_{1}^{2}+\sin^{2}\theta_{1}d\phi_{1}^{2})+\frac{1}{4}\sin^{2}\xi(d\theta_{2}^{2}+\sin^{2}\theta_{2}d\phi_{2}^{2})].
     \end{split}
 \end{equation}
Now let us define the $X_{i}$'s in terms of the polar coordinates as follows :
\begin{equation}
  W_{1}=  X_{1}+iX_{2}=Rr_{1}e^{i\Phi_{1}}, W_{2}=X_{3}+iX_{4}=Rr_{2}e^{i\Phi_{2}}
  \label{constraint3}
\end{equation}
\begin{equation}
    W_{3}=X_{5}+iX_{6}=Rr_{3}e^{i\Phi_{3}}, W_{4}=X_{7}+iX_{8}=Rr_{4}e^{i\Phi_{4}}
    \label{constraint4}
\end{equation}Therefore the embedding of the string in $R_{t}\times \mathbb{CP}^{3}$ may be reduced as 
 \begin{equation}
 z=Z(\tau,\sigma)=\frac{R}{2}e^{it(\tau,\sigma)} , w_{a}=W_{a}(\tau,\sigma)=Rr_{a}(\tau,\sigma)e^{i\Phi_{a}(\tau,\sigma)}.
 \label{parameter}
 \end{equation}In the case of embedding in $\mathbb{CP}^{3}$, it must be :
 \begin{equation}
 \sum_{a=1}^{4}W_{a}\bar{W_{a}}=R^{2}, \sum_{a=1}^{4}(W_{a}\partial_{\alpha}\bar{W_{a}}-\bar{W_{a}}\partial_{\alpha}W_{a})=0
 \label{constraint5}
 \end{equation}In terms of the embedding coordinates, the $\mathbb{CP}^{3}$ constraints become
 \begin{equation}
 \sum_{a=1}^{4}{r_{a}}^{2}(\tau,\sigma)=0,\sum_{a=1}^{4}{r_{a}}^{2}(\tau,\sigma)\partial_{\alpha}\Phi_{a}(\tau,\sigma)=0
 \label{constraints1}
 \end{equation}For this embedding, the metric induced on the string worldsheet $G_{\alpha \beta}$ and the $B_{\alpha \beta}$ is given by,
 \begin{equation}
  G_{\alpha\beta}=-\partial_{(\alpha}Z_{\beta)}\bar{Z}+\sum_{a=1}^{4}\partial_{(\alpha}W_{a}\partial_{\beta)}\bar{W_{a}},
  \end{equation}
  \begin{equation}
  B_{\alpha\beta} = B_{MN}\partial_\alpha X^M\partial_\beta X^N=\frac{b}{2}\epsilon^{\alpha\beta}\left[\partial_{(\alpha}Z\partial_{\beta)}\bar{Z}+\sum_{a=1}^{4}\partial_{(\alpha}W_{a}\partial_{\beta)}\bar{W_{a}}\right] \ .
  \end{equation}Putting the expressions of $Z$ and $W_{a}$ we get
  \begin{equation}
 G_{\alpha\beta}=-\frac{R^{2}}{4}(\partial_{\alpha}t\partial_{\beta}t)+R^{2}\sum_{a=1}^{4}(\partial_{\alpha}r_{a}\partial_{\beta}r_{a}+{r_{a}}^{2}\partial_{\alpha}\Phi_{a}\partial_{\beta}\Phi_{a}),
 \end{equation}\begin{equation}
   B_{\alpha\beta} = B_{MN}\partial_\alpha X^M\partial_\beta X^N= b R^{2}\sum_{a=1}^{4}r_{a}\left[\partial_{\sigma}r_{a}\partial_{\tau}\Phi_{a}-\partial_{\tau}r_{a}\partial_{\sigma}\Phi_{a}\right] \ .
 \end{equation}
 The corresponding Lagrangian in target space now becomes
 \begin{eqnarray}
 \mathcal{L} = -\sqrt{2\lambda}\big\{\frac{(\partial_{\tau}t)^{2}-(\partial_{\sigma}t)^{2}}{4}+\sum_{a=1}^{4}(\partial_{\sigma}r_{a})^{2} -\sum_{a=1}^{4}
   (\partial_{\tau}r_{a})^{2} \nonumber \\
    -\sum_{a=1}^{4}{r_{a}}^{2}[(\partial_{\tau}
    \Phi_{a})^{2}-(\partial_{\sigma}\Phi_
   {a})^{2}]\big\}
   -{b}\sqrt{2\lambda}\sum_{a=1}^{4}r_{a}(\partial_{\sigma}r_{a}\partial_{\tau}\Phi_{a}-\partial_{\tau}r_{a}\partial_{\sigma}\Phi_{a})\nonumber \\ 
+ \sqrt{8\lambda}\Lambda(\sum_{a=1}^{4}{r_{a}}^{2}-1)
  +\sqrt{8\lambda}\Lambda_{0}\sum_{a=1}^{4}({r_{a}}^{2}\partial_{\tau}\Phi_{a})+\sqrt{8\lambda}\Lambda_{1}\sum_{a=1}^{4}({r_{a}}^{2}\partial_{\sigma}\Phi_{a}).
   \label{lagrangian4}
   \end{eqnarray}
   Here $M,N=\tau,\sigma$ and $\Lambda ,\Lambda_{0},\Lambda_{1}$ are suitable Lagrange multipliers corresponding to the constraints.
   
\subsection{Lagrangian and Hamiltonian formulation}
NR system, being an integrable modification of the first proposed Neumann integrable model, illustrates the constrained motion of a harmonic oscillator of unit mass on a $(N-1)$ dimensional unit sphere under another centrifugal potential barrier. The Lagrangian for such a system is given by
\begin{eqnarray}
L=\frac{1}{2}\sum_{i=1}^{N}\left[x^{'2}_{i}+x^{2}_{i}\left(K^{'2}_{i}-\omega^{2}_{i}\right)\right]-\frac{\Lambda}{2}\left(\sum_{i=1}^{N}x_{i}^{2}-1\right),
\label{lagrangian}
\end{eqnarray}where $K^{'}_{i}=\frac{v^2_{i}}{x_{i}^{2}}$, with $v^2_i$ being a constant 
and $\Lambda$ is a suitable Lagrange multiplier to deal with the spherical geometry.
The corresponding equation of motion is 
\begin{equation}
x_{i}^{''}=\left(K_{i}^{'2}-\omega_{i}^{2}+\Lambda\right)x_{i}.
\end{equation}The Hamiltonian for such a system may be written as
\begin{equation}
H=\frac{1}{2}\sum_{i=1}^{2}\left[x_{i}^{'2}-x_{i}^{2}\left(K_{i}^{'2}-\omega_{i}^{2}\right)\right],
\label{hamiltonian}
\end{equation}where $\sum_{i=1}^{N}x_{i}^{2}=1$.
 We wish to study the spinning string in $R_{t}\times \mathbb{CP}^{3}$ background in the presence of $B_{NS}$ holonomy. We use the following parametrization: 
\begin{eqnarray}
t=\kappa\tau, \>\> r_{a}(\tau,\sigma)=r_{a}(\zeta), \>\> \Phi_{a}(\tau,\sigma)=\omega_{a}\tau+f_{a}(\zeta), \>\> \zeta=\alpha\sigma+\beta\tau \ ,
\label{ansatz}
 \end{eqnarray}where  $\kappa ,\omega_{a} , \alpha , \beta$ are constants. Using the ansatz (\ref{ansatz}) for the string rotating in $R_{t} \times \mathbb{CP}^{3}$, the Lagrangian (\ref{lagrangian4}) becomes,
\begin{eqnarray}
 \mathcal{L}= -\sqrt{2\lambda}\left[\frac{\kappa^{2}}{4}+(\alpha^{2}-\beta^{2})
 \sum_{a=1}^{4}\left({r_{a}}^{'2}+{r_{a}}^{2}(f_{a}^{'}-
 \frac{\beta\omega_{a}}{\alpha^{2}-\beta^{2}})^{2}-\frac{\alpha^{2}
 \omega_{a}^{2}r_{a}^{2}}{(\alpha^{2}-\beta^{2})^{2}}\right)\right] \nonumber \\ 
 -b\sqrt{2\lambda}\sum_{a=1}^{4}r_{a}^{'}r_{a}\alpha\omega_{a}+\sqrt{8\lambda}
 \Lambda(\sum_{a=1}^{4}{r_{a}}^{2}-1)
  +\sqrt{8\lambda}\Lambda_{0}\sum_{a=1}^{4}({r_{a}}^{2}\omega_{a})
 +\sqrt{8\lambda}\Lambda_{1}\sum_{a=1}^{4}({r_{a}}^{2}f_{a}^{'}). 
 \nonumber \\
 \label{lagrangian1}
 \end{eqnarray}
   Equation of motion for $f_{a}$ is
   \begin{equation}
   f_{a}^{'}=\frac{1}{\alpha^{2}-\beta^{2}}\left[\frac{C_{a}}{r_{a}^{2}}+\beta\omega_{a}+\Lambda_{1}\right],
   \end{equation}  where $C_{a}$'s are proper integration constants.
  Putting this expression of $f_{a}$ in the Lagrangian (\ref{lagrangian1})we get 
   \begin{equation}
   \begin{split}
   \mathcal{L}=&-\sqrt{2\lambda}\left[\frac{\kappa^{2}}{4}+(\alpha^{2}-\beta^{2})\sum_{a=1}^{4}\left(r_{a}^{'^2}+\frac{1}{(\alpha^{2}-\beta^{2})}(\frac{C_{a}^{2}}{r_{a}^{2}}+2C_{a}\Lambda_{1}+\Lambda_{1}^{2}r_{a}^{2})-\frac{\alpha^{2}\omega_{a}^{2}r_{a}^{2}}{(\alpha^{2}-\beta^{2})^{2}}\right)\right]\\&-B\sqrt{2\lambda}\sum_{a=1}^{4}\alpha\omega_{a}r_{a}^{'}r_{a}+\sqrt{8\lambda}\Lambda(\sum_{a=1}^{4}{r_{a}}^{2}-1)
   +\sqrt{8\lambda}\Lambda_{0}\sum_{a=1}^{4}({r_{a}}^{2}\omega_{a})\\&
   +\sqrt{8\lambda}\Lambda_{1}\sum_{a=1}^{4}\frac{1}{(\alpha^{2}-\beta^{2})}(C_{a}+\beta\omega_{a}r_{a}^{2}+\Lambda_{1}r_{a}^{2}).
   \end{split}
   \label{lagrangian2}
   \end{equation}From (\ref{lagrangian2}), we calculate the equation of motion for $r_{a}$ as 
   \begin{equation}
   (\alpha^{2}-\beta^{2})r_{a}^{''}-\frac{C_{a}}{(\alpha^{2}-\beta^{2})r_{a}^{3}}+[2(\Lambda+\Lambda_{0}\omega_{a})+\omega_{a}^{2}+\frac{(\Lambda_{1}+\beta\omega_{a})}{(\alpha^{2}-\beta^{2})}]r_{a}=0.
   \label{EOM}
   \end{equation}    
   We note that, this equation can also be derived from the following Lagrangian:
   \begin{equation}
       \begin{split}
   \mathcal{L}= &\sum_{a=1}^{4}[(\alpha^{2}-\beta^{2})r_{a}^{'2}-\frac{1}{(\alpha^{2}-\beta^{2})}\frac{C_{a}}{r_{a}^{2}}-\omega_{a}^{2}r_{a}^{2}]+\sum_{a=1}^{4}\alpha\omega_{a}r_{a}r_{a}^{'}-2\Lambda(\sum_{a=1}^{4}{r_{a}}^{2}-1)\\&
   -2\Lambda_{0}\sum_{a=1}^{4}({r_{a}}^{2}\omega_{a})+\sum_{a=1}^{4}\frac{1}{(\alpha^{2}-\beta^{2})}(\beta\omega_{a}r_{a}^{2}+\Lambda_{1}r_{a}^{2}).
   \end{split}
   \label{lagrangian3}
   \end{equation}Here it is quite obvious from equation (\ref{lagrangian3}) that the Lagrangian is that for the NR system only with an extra term added due to the presence of $B_{NS}$ two-form holonomy through $\mathbb{CP}^{1}$ and two additional constraints (\ref{constraints1}) to deal with the geometry of $\mathbb{CP}^{3}$. Now the equation of motion for $\Lambda_{1}$ gives
   \begin{center}
   $\sum_{a=1}^{4}(\beta\omega_{a}+\Lambda_{1})=0$,\\
   \end{center}so that the term $\sum_{a=1}^{4}\frac{1}{(\alpha^{2}-\beta^{2})}(\beta\omega_{a}r_{a}^{2}+\Lambda_{1}r_{a}^{2})$ in the Lagrangian becomes zero. Now from the constraints of $AdS_{4}\times \mathbb{CP}^{3}$, we get the $C_{a}$'s as
   \begin{equation}
   \Lambda_{1}=-\sum_{a=1}^{4}C_{a}=0,~~C_{a}^{2}=\beta^{2}\omega_{a}^{2}r_{a}^{4}.
   \end{equation}
 The Hamiltonian for such system is given as
   \begin{equation}
      H_{NR}= (\alpha^{2}-\beta^{2})\sum_{a=1}^{4}\left[r_{a}^{'2}+\frac{1}{(\alpha^{2}-\beta^{2})^{2}}(\frac{{C_{a}}^{2}}{{r_{a}}^{2}}+\alpha^{2}{\omega_{a}}^{2}{r_{a}}^{2})\right]=\frac{\alpha^{2}+\beta^{2}}{\alpha^{2}-\beta^{2}}\frac{\kappa^{2}}{4},
     \label{Hamiltonian}
   \end{equation} whose form is exactly the same as equation (\ref{hamiltonian}), thereby supporting the NR approach of studying the gravity dual of ABJ theory in planar limit. We note that in deriving the above relations, we use the Virasoro constraints, 
  $G_{\tau\tau}+G_{\sigma\sigma}=0$ and $G_{\tau\sigma}=0$ which gives the conserved Hamiltonian $H_{NR}$ and the relation between the embedding coordinates and the arbitrary constants $C_{a}$.
 Another relation $\sum_{a=1}^{4}\omega_{a}C_{a}+\frac{\beta\kappa^{2}}{4}=0$ along with the Hamiltonian helps to satisfy both the Virasoro constraints simultaneusly. For closed strings, $ r_{a}$ and $f_{a}$ satisfy the periodicity conditions as,
  \begin{equation}
  r_{a}(\zeta+2\pi\alpha)=r_{a}(\zeta), f_{a}(\zeta+2\pi\alpha)=f_{a}(\zeta)+2 \pi  n_{a},
  \end{equation}where $n_{a}$ is the integer winding number.
  
\subsection{Integrals of motion}
Integrability of any system requires the existence of infinite number of conserved quantities, also known as the integrals of motion, to be in involution. K. Uhlenback fist introduced the integrals of motion for Neumann model which states that there must be $N$ number of integrals of motion $I_{i}$ such that \cite{Uhlenbeck}
\begin{equation}
\left\{I_{i},I_{j}\right\} =0 ~  \forall  ~i,j \in \left\{1,2,......,N\right\} \ ,
\end{equation} so that the integrable features of NR system exists. For any arbitrary values of the constants $v^2_{i}$, the integrals of motion for the NR system assume the following form
\begin{equation}
I_{i}=x_{i}^{2}+\sum_{j\neq i}\frac{1}{\omega_{i}^{2}-\omega_{j}^{2}}\left[\left(x_{i}x_{j}^{'}-x_{j}x_{i}^{'}\right)^{2}+v_{i}^{2}\frac{x_{j}^{2}}{x_{i}^{2}}+v_{j}^{2}\frac{x_{i}^{2}}{x_{j}^{2}}\right] \ .
\end{equation}
To construct the integrals of motion for the string moving in ABJ background we use the parametrization (\ref{parameter}). With this, the Lagrangian in the desired background reads as\footnote{here we consider $b=1$ for simplicity}
\begin{equation}
\begin{split}
\mathcal{L}=&-(\partial_{\tau}t)^{2}+(\partial_{\sigma}t)^{2}+\sum_{a}\left(\partial_{\tau}W_{a}\partial_{\tau}\bar{W}_{a}-\partial_{\sigma}W_{a}\partial_{\sigma}\bar{W}_{a}\right)\\ &
+\frac{i}{2 } \sum_{a}\left(-\partial_{\tau}W_{a}\partial_{\sigma}\bar{W}_{a}+\partial_{\tau}\bar{W}_{a}\partial_{\sigma}W_{a}\right)
-\Lambda\left(\sum_{a}W_{a}\bar{W}_{a}-1\right)\\&-\Lambda_{0}\sum_{a}\left(W_{a}\partial_{\tau}\bar{W}_{a}-\bar{W}_{a}\partial_{\tau}W_{a}\right)-\Lambda_{1}\sum_{a}\left(W_{a}\partial_{\sigma}\bar{W}_{a}-\bar{W}_{a}\partial_{\sigma}W_{a}\right).
\end{split}
\end{equation}For convenience, let us use the following ansatz\cite{Kruczenski:2006pk}
\begin{equation}
    W_{a}=x_{a}(\zeta)e^{i\omega_{a}\tau},
    \label{5.6}
    \end{equation}where $\zeta=\alpha\sigma+\beta\tau$ and $x_{a}(\zeta)=r_{a}(\zeta)e^{if_{a}(\zeta)}$.\\
    The equation of motion for $x_{a}$ is given by
    \begin{equation}
        (\alpha^{2}-\beta^{2})x_{a}^{''}-(2i\beta\omega_{a}-2\Lambda_{0}\beta-2\Lambda_{1}\alpha)x_{a}^{'}-(\omega_{a}^{2}+2i\Lambda_{0}\omega_{a}-\Lambda)x_{a}=0.
    \end{equation}
    It may be noted that this equation of motion can also be derived from the following Lagrangian
    \begin{equation}
    \begin{split}
    \mathcal{L}=&\left[\sum_{a}(\alpha^{2}-\beta^{2})x_{a}^{'}\bar{x}_{a}^{'}+i\beta\sum_{a}\omega_{a}(x_{a}^{'}\bar{x}_{a}-\bar{x}_{a}^{'}x_{a})-\sum_{a}\omega_{a}^{2}x_{a}\bar{x}_{a}\right]\\& +\frac{1}{2}\sum_{a}\alpha\omega_{a}(x_{a}\bar{x}_{a}^{'}+\bar{x}_{a}x_{a}^{'})+\Lambda\left(\sum_{a}x_{a}\bar{x}_{a}-1\right)+\Lambda_{0}\beta\sum_{a}\left(x_{a}\bar{x}_{a}^{'}-\bar{x}_{a}x_{a}^{'}\right)\\& -2i\Lambda_{0}\sum_{a}\omega_{a}x_{a}\bar{x}_{a}+\Lambda_{1}\alpha\sum_{a}\left(x_{a}\bar{x}_{a}^{'}-\bar{x}_{a}x_{a}^{'}\right).
    \end{split}
    \end{equation}This Lagrangian is equivalent to (\ref{lagrangian3}). 
    The momenta $p_{a}$ conjugate to $x_{a}$ can be derived as
   \begin{equation}
       p_{a}=\frac{\partial\mathcal{L}}{\partial\bar{x}_{a}^{'}}=(\alpha^{2}-\beta^{2})x_{a}^{'}-i\beta\omega_{a}x_{a}+\frac{\alpha}{2}\omega_{a}x_{a}+\Lambda_{0}\beta x_{a}+\Lambda_{1}\alpha x_{a},
   \end{equation} 
    Therefore,
    \begin{equation}
        (\bar {x}_{b} p_{a}-x_{a}\bar {p}_{b})=(\alpha^{2}-\beta^{2})(x_{a}^{'}\bar{x}_{b}-x_{a}\bar {x}_{b}^{'})-i\beta(\omega_{a}+\omega_{b})x_{a}\bar{x}_{b}+\frac{\alpha}{2}(\omega_{a}+\omega_{b})x_{a}\bar{x}_{b}.
        \label{IOM1}
    \end{equation} Now, we know that the form of the integral of motion for any NR system is 
   \begin{equation}
       F_{\alpha}=\alpha^{2}x_{a}x_{a}^{'}+\sum_{b\neq a}\frac{|\bar{x}_{b}p_{a}-x_{a}\bar {p}_{b}|^{2}}{(\omega_{a}^{2}-\omega_{b}^{2})}.
       \label{IOM2}
   \end{equation}
  Using equation (\ref{5.6}), we get,
  \begin{equation}
      x_{a}^{'}=(r_{a}^{'}+ir_{a}f_{a}^{'})e^{if_{a}}.
      \label{IOM3}
  \end{equation}Again, from the equations of motion we get
  \begin{equation}
 i (\alpha^{2}-\beta^{2})(x_{a}^{'}\bar{x}_{a}-\bar{x}_{a}^{'}x_{a})^{'}=   [-2\beta\omega_{a}-2i(\Lambda_{0}\beta+\Lambda_{1}\alpha)]x_{a}\bar{x}_{a} .
  \label{IOM5}
  \end{equation}Now using equations (\ref{IOM1}-\ref{IOM5}) we get,
 \begin{align*}
 \alpha^{2}x_{a}\bar{x}_{a}^{'}&=\alpha^{2}r_{a}^{2},\\
     |\bar{x}_{b}p_{a}-x_{a}\bar {p}_{b}|^{2}&=(\bar{x}_{b}p_{a}-x_{a}\bar {p}_{b})(x_{b}\bar{p}_{a}-\bar{x}_{a}p_{b})\\
     &=(\alpha^{2}-\beta^{2})(r_{a}^{'}r_{b}-r_{a}r_{b}^{'})^{2}+(\frac{C_{a}r_{b}}{r_{a}}+\frac{C_{b}r_{a}}{r_{b}})^{2}+\\ & 2\beta\left[C_{a}r_{b}^{2}\omega_{a}+C_{a}r_{b}^{2}\omega_{b}+C_{b}r_{a}^{2}\omega_{a}+C_{b}r_{a}^{2}\omega_{b}\right]+\frac{\alpha^{2}}{4}(\omega_{a}+\omega_{b})^{2}r_{a}^{2}r_{b}^{2} .
 \end{align*}
This finally yields 
 \begin{eqnarray}
F_{\alpha} &=& \alpha^{2}r_{a}^{2}+\sum_{b\neq a}\frac{(r_{a}^{'}r_{b}-r_{a}r_{b}^{'})^{2}}{(\omega_{a}^{2}-\omega_{b}^{2})}+\sum_{b\neq a}\frac{1}{(\omega_{a}^{2}-\omega_{b}^{2})}\left(\frac{C_{a}r_{b}}{r_{a}}+\frac{C_{b}r_{a}}{r_{b}}\right)^{2} \nonumber \\
&+& \frac{\alpha^{2}}{4}\sum_{b\neq a}\left(\frac{\omega_{a}+\omega_{b}}{\omega_{a}-\omega_{b}}\right)r_{a}^{2}r_{b}^{2} .
\end{eqnarray} These are the Uhlenback integrals of motion for the closed rotating string in $R_t \times \mathbb{CP}^{3}$ in the presence of the $B_{NS}$ holonomy.
\section{Dyonic magnon on $R_t\times\mathbb{CP}^3$ with flux}
 Let us assume that the string is spinning in the given background with two independent angular momenta $\omega_{1} = -\omega_{3}$ and $\omega_{2} = -\omega_{4}$. For such a system we consider the embedding coordinates as
  $r_{1}=r_{3}=\frac{1}{\sqrt{2}}\sin\theta$ and $r_{2}=r_{4}=\frac{1}{\sqrt{2}}\cos\theta.$
The equation of motion (\ref{EOM}) then reduces to 
  \begin{equation}
 \theta^{'2}(\zeta)=\frac{1}{(\alpha^{2}-\beta^{2})^{2}}[(\alpha^{2}+\beta^{2})\frac{\kappa^{2}}{4}-\frac{2(C_{1}^{2}+C_{3}^{2})}{(\sin\theta)^{2}}-\frac{2(C_{2}^{2}+C_{4}^{2})}{(\cos\theta)^{2}}-\alpha^{2}(\omega_{1}^{2}(\sin\theta)^{2}+\omega_{2}^{2}(\cos\theta)^{2})],
 \label{theta}
   \end{equation} where we have used the Hamiltonian(\ref{Hamiltonian}). We wish to study the giant magnon solution. For this we put $C_{2}=C_{4}=0$ so that $\sum_{a=1}^{4}C_{a}\omega_{a}=-\frac{\beta\kappa^{2}}{4}$ yields  $C_{1}=-C_{3}=-\frac{\beta\kappa^{2}}{8\omega_{1}}$. Thus the expression for (\ref{theta})  reduces to
 \begin{equation}
 \begin{split}
  (\cos\theta)^{'}=&\mp\frac{1}{(\alpha^{2}-\beta^{2})}[(\alpha^{2}+\beta^{2})\frac{\kappa^{2}}{4}-(\alpha^{2}+\beta^{2})\frac{\kappa^{2}}{4}\cos^{2}\theta-\frac{\beta^{2}\kappa^{4}}{16\omega_{1}^{2}}\\&-\alpha^{2}\omega_{1}^{2}+2\alpha^{2}\omega_{1}^{2}\cos^{2}\theta-\alpha^{2}\omega_{2}^{2}\cos^{2}\theta+\alpha^{2}(\omega_{2}^{2}-\omega_{1}^{2})\cos^{4}\theta]^{\frac{1}{2}},
  \end{split}
  \end{equation}with a solution
\begin{equation}
\cos\theta=z_{+}dn(C\zeta|m),
\end{equation}where 
  \begin{equation}
  {z_{\pm}}^{2}=\frac{1}{2(1-\frac{{\omega_{2}}^{2}}{{\omega_{1}}^{2}})}\left[y_{1}+y_{2}-\frac{{\omega_{2}}^{2}}{{\omega_{1}}^{2}}\pm\sqrt{({y_{1}}-{y_{2}})^{2}-[2(y_{1}+y_{2}-2y_{1}y_{2})-\frac{{\omega_{2}}^{2}}{{\omega_{1}}^{2}}]\frac{{\omega_{2}}^{2}}{{\omega_{1}}^{2}}}\right] \ .
  \end{equation} In the above exprerssion, $y_{1}=1-\frac{\kappa^{2}}{4{\omega_{1}}^{2}}$ and $y_{2}=1-\frac{\beta^{2}\kappa^{2}}{4\alpha^{2}{\omega_{1}}^{2}};$ $C=\mp\frac{\alpha\sqrt{(\omega_{1}^{2}-\omega_{2}^{2})}}{(\alpha^{2}-\beta^{2})}z_{+}$ and
  $m=1-\frac{z_{-}^{2}}{z_{+}^{2}}$.\\
For $f_{a}=\frac{1}{\alpha^{2}-\beta^{2}}\int d\zeta (\frac{C_{a}}{r_{a}^{2}}+\beta\omega_{a})$, one has
 \begin{equation}
 f_{1}=-f_{3}=\mp\frac{\beta}{\alpha z_{+}\sqrt{1-\frac{\omega^{2}_{2}}{\omega_{1}^{2}}}}\left[\boldsymbol{K}(1-\frac{z_{-}^{2}}{z_{+}^{2}})-\frac{\kappa^{2}}{4\omega_{1}^{2}(1-z_{+}^{2})}\big\{\Pi(am(C\zeta),-\frac{(z_{+}^{2}-z_{-}^{2})}{(1-z_{+}^{
 2})}|m)\big\}\right],
 \end{equation}
 \begin{equation}
 f_{2}=-f_{4}=\mp\frac{\beta\omega_{2}}{\alpha\omega_{1}z_{+}}\frac{\boldsymbol{K}(1-\frac{z_{-}^{2}}{z_{+}^{2}})}{\sqrt{1-\frac{\omega_{2}^{2}}{\omega_{1}^{2}}}}.
 \end{equation}
The full string solution in this case using NR integrable system is the same as the ones obtained in \cite{Ahn:2008hj} for the string in $AdS_4 \times \mathbb{CP}^3$ without the $B_{NS}$ flux. 
 
\subsection{Dispersion relation and finite size effect}
As the Lagrangian does not depend on $t$ and $\phi_{a}$ we have the conserved charges as\\
\begin{equation}
E=-\int{d\sigma \frac{\partial\mathcal{L}}{\partial(\partial_{\tau}t)}},~~ J_{a}=\int{d \sigma\frac{\partial\mathcal{L}}{\partial(\partial_{\tau}\phi_{a})}},
\end{equation}with $a=1,2$.
Therefore,
\begin{equation}
E_{S}=\frac{\kappa \sqrt{2\lambda}}{2\alpha}\int d\zeta ,\>\>   J_{a}=2\frac{\sqrt{2\lambda}}{(\alpha^{2}-\beta^{2})}\int d\zeta (\frac{\beta}{\alpha}C_{a}+\alpha^{2}\omega_{a}r_{a}^{2})-\frac{b}{4\pi}\int r_{a}r_{a}^{'}d\zeta
\end{equation}
In what follows we will be looking at the conserved charges for $\alpha^{2}>\beta^{2}$. $E$ and $J_{a}$ become
\begin{equation}
\mathcal{E}=\frac{E_{s}}{\sqrt{2\lambda}}=\frac{\kappa}{2\alpha} \int d\zeta = \frac{2\kappa(1-\frac{\beta^{2}}{\alpha^{2}})}{\omega_{1}z_{+}\sqrt{1-\frac{\omega_{2}^{2}}{\omega_{1}^{2}}}}\boldsymbol{K}(1-\frac{z_{-}^{2}}{z_{+}^{2}}).
\label{energy}
\end{equation}
\begin{equation}
\mathcal{J}_{1}=\frac{J_{1}}{\sqrt{2\lambda}}=\frac{2}{\alpha^{2}-\beta^{2}} \int (\frac{\beta}{\alpha}C_{a}+\alpha\omega_{1}r_{1}^{2})d\zeta -\frac{b}{4\pi\sqrt{2\lambda}} \int r_{1}r_{1}^{'}d\zeta
\end{equation}\begin{equation}
\mathcal{J}_{2} =\frac{2}{\alpha^{2}-\beta^{2}} \int \alpha\omega_{2}r_{2}^{2}d\zeta +\frac{b}{8\pi\sqrt{2\lambda}} \int r_{2}r_{2}^{'} d\zeta
\end{equation}Therefore, the final expression of the currents after doing the integration are as follows:
\begin{equation}
\mathcal{J}_{1} =\frac{2}{z_{+}^{2}\sqrt{1-\frac{\omega_{2}^{2}}{\omega_{1}^{2}}}}z_{+}(1-\frac{\beta^{2}\kappa^{2}}{4\alpha^{2}\omega_{1}^{2}})\boldsymbol{K}(1-\frac{z_{-}^{2}}{z_{+}^{2}})-\frac{2}{z_{+}^{2}\sqrt{1-\frac{\omega_{2}^{2}}{\omega_{1}^{2}}}}\boldsymbol{E}(1-\frac{z_{-}^{2}}{z_{+}^{2}})-\frac{b}{4} \left(z_{+}^{2}-z_{-}^{2}\right) \ ,
\label{current1}
\end{equation}\begin{equation}
\mathcal{J}_{2}=\frac{2}{1-\frac{\omega_{2}^{2}}{\omega_{1}^{2}}}\frac{\omega_{2}}{\omega_{1}}z_{+}\textbf{E}\left(1-\frac{z_{-}^{2}}{z_{+}^{2}}\right)+\frac{b}{4} \left(z_{+}^{2}-z_{-}^{2}\right) \ ,
\label{current2}
\end{equation}
\begin{equation}
\mathcal{J}_{3}=-\mathcal{J}_{1}-\frac{b}{2}\left(z_{+}^{2}-z_{-}^{2}\right) \ ,
\end{equation}
\begin{equation}
\mathcal{J}_{4}= -\mathcal{J}_{3}+\frac{b}{2}\left(z_{+}^{2}-z_{-}^{2}\right) \ .
\end{equation}
Therefore it is quite obvious that the constraints of $\mathbb{CP}^{3}$ geometry support
\begin{equation}
\sum_{a=1}^{4}\mathcal{J}_{a}=0
\end{equation}by using the NR approach. The computation of $\Delta\Phi_{1}$ gives 
 \begin{equation}
 \begin{split}
 p=&\Delta\Phi_{1}=2 \int_{\theta_{min}}^{\theta_{max}}\frac{d\theta}{\theta^{'}}f_{1}^{'}\\&= \frac{2\beta}{\alpha z_{+}\sqrt{1-\frac{\omega^{2}_{2}}{\omega_{1}^{2}}}}\left[\boldsymbol{K}(1-\frac{z_{-}^{2}}{z_{+}^{2}})-
 \frac{\kappa^{2}}{4\omega_{1}^{2}(1-z_{+}^{2})}\big\{\boldsymbol{\Pi}(am(C\zeta),
 -\frac{(z_{+}^{2}-z_{-}^{2})}{(1-z_{+}^{
 2})}|m)\big\}\right] .
 \end{split}
 \end{equation}
Here, let us assume, $u\equiv \frac{\omega_{2}^{2}}{\omega_{1}^{2}}$, $v\equiv \frac{\beta}{\alpha}$, $\epsilon\equiv \frac{z_{-}^{2}}{z_{+}^{2}}$ , where $u$ and $v$ are also functions of $\epsilon$. 
 Considering $\frac{z_{-}^{2}}{z_{+}^{2}}=\epsilon\rightarrow 0$ the leading order  value  of $z_{+}$ is
\begin{equation}
z_{+}=\sqrt{\mathcal{J}_{2}^{2}+4\sin^{2}\frac{p}{2}},
\end{equation}and hence


\begin{equation}
\epsilon=16 \exp\left[-2\frac{(\mathcal{J}_{1}+\mathcal{J}_{2}+\sqrt{\mathcal{J}_{2}^{2}+4\sin^{2}\frac{p}{2}})}{\mathcal{J}_{2}^{2}+4\sin^{4}\frac{p}{2}}\sqrt{\mathcal{J}_{2}^{2}+4\sin^{2}\frac{p}{2}}\sin^{2}\frac{p}{2}\right].
\end{equation}Putting all the expansions of the elliptic integrals and their coefficients we get the dispersion relation as 
\begin{equation}
\begin{split}
\mathcal{E}-\mathcal{J}_{1}=&\sqrt{\mathcal{J}_{2}^{2}+4\sin^{2}\frac{p}
{2}}- \frac{32\sin^{4}\frac{p}{2}}{\sqrt{\mathcal{J}_{2}^{2}+4\sin^{2}\frac{p}{2}}}\\&\exp\left[-2\frac{(\mathcal{J}_{1}+\mathcal{J}_{2}+\sqrt{\mathcal{J}_{2}^{2}+4\sin^{2}\frac{p}{2}})}{\mathcal{J}_{2}^{2}+4\sin^{4}\frac{p}{2}}\sqrt{\mathcal{J}_{2}^{2}+4\sin^{2}\frac{p}{2}}\sin^{2}\frac{p}{2}\right]\\&-\frac{b}{4}\left(\mathcal{J}_{2}^{2}+4\sin^{2}\frac{p}{2}\right)\\&\left(\exp\left[-2\frac{(\mathcal{J}_{1}+\mathcal{J}_{2}+\sqrt{\mathcal{J}_{2}^{2}+4\sin^{2}\frac{p}{2}})}{\mathcal{J}_{2}^{2}+4\sin^{4}\frac{p}{2}}\sqrt{\mathcal{J}_{2}^{2}+4\sin^{2}\frac{p}{2}}\sin^{2}\frac{p}{2}\right]-1\right).
\end{split}
\label{finitesizecorrection}
\end{equation}This result resembles with the dispersion relation for dyonic giant magnon in the $R_{t}\times S^{3}$  \cite{Chen:2006gea,Hatsuda:2008gd} with an extra $B$ dependent term. For the string rotating in $R_{t}\times \mathbb{CP}^{3}$ subspace there exists a similar dispersion relation between $\mathcal{E}$, $\mathcal{J}_{3}$  and $\mathcal{J}_{4}$ with a form exactly the same as that of (\ref{finitesizecorrection}). Again with one angular momentum $\mathcal{J}_{2}$ to be zero it reduces to the giant magnon dispersion relation in $R_{t}\times S^{2}$ subspace \cite{Jain:2008mt}.
\section{Pulsating string in dual background of ABJ Model}
In this section, we analyze the case of closed pulsating string in $R_{t} \times \mathbb{CP}^{3}$ with $B_{NS}$ flux. It will be worth showing that the pulsating strings with two-form NS-NS fluxes in $R_t \times \mathbb{CP}^{3}$ can also be reduced to a deformed NR system. The embedding ansatz for closed pulsating string in $R_{t}\times\mathbb{CP}^{3}$ is \cite{Hernandez:2018gcd}:\\
\begin{subequations}
\begin{align}
Z(\tau)&=Y_{0}+iY_{3}=\frac{R}{2}z_{0}(\tau)e^{ih_{0}(\tau)}\\
 W_{1}(\tau,\sigma)&=X_{1}+iX_{2}=Rr_{1}(\tau)e^{i(f_{1}(\tau)+m_{1}\sigma)}\\
 W_{2}(\tau,\sigma)&=X_{3}+iX_{4}=Rr_{2}(\tau)e^{i(f_{2}(\tau)+m_{2}\sigma)}\\
 W_{3}(\tau,\sigma)&=X_{5}+iX_{6}=Rr_{2}(\tau)e^{i(f_{3}(\tau)+m_{3}\sigma)}\\
  W_{4}(\tau,\sigma)&=X_{7}+iX_{8}=Rr_{4}(\tau)e^{i(f_{4}(\tau)+m_{4}\sigma)}
\end{align}
\end{subequations}where, $z_{0}=z_{0}(\tau)$ and $r_{a}=r_{a}(\tau)$, with $a=1,2,3,4,$. The winding numbers $m_{a}$ are kept only along the $\sigma$ direction to make the time direction single-valued.
Taking such an ansatz and considering all the constraints, the Lagrangian for the pulsating string may be derived as:
\begin{eqnarray}
\mathcal{L}=-\sqrt{2\lambda}\left[\frac{\dot{z}_{0}^{2}+z_{0}^{2}\dot{h}_{0}^{2}}{4}\right]+\sqrt{2\lambda}\sum_{a=1}^{4}\left[\dot{r}_{a}^{2}+r_{a}^{2}\dot{f}_{a}^{2}-r_{a}^{2}m_{a}^{2}\right]\nonumber\\+b\sqrt{2\lambda}\sum_{a=1}^{4}r_{a}\dot{r}_{a}m_{a}-\frac{\Lambda}{2}\sqrt{8\lambda}\left(\sum_{a=1}^{4}r_{a}^{2}-1\right)-\frac{\Tilde{\Lambda}}{2}\sqrt{8\lambda}\left(z_{0}^{2}+1 \right)\nonumber\\
-\sqrt{8\lambda}\frac{\Lambda_{1}}{2}\sum_{a=1}^{4}r_{a}^{2}\dot{f}_{a}-\sqrt{8\Lambda}\frac{\Lambda_{2}}{2}\sum_{a=1}^{4}r_{a}^{2}m_{a},
\label{Pulsating Lagrangian}
\end{eqnarray}where the derivative with respect to $\tau$ is denoted by dots. $\Lambda$,$\tilde{\Lambda}$,$\Lambda_{1}$ and $\Lambda_{2}$ are suitable Lagrange multipliers. The equations of motion for $z_{0}$ and $f_{a}$ are given by
\begin{equation}
    \ddot{z_{0}}-\frac{C_{0}^{2}}{z_{0}^{3}}+4\Tilde{\Lambda}z_{0}=0,
    \label{z equation}
\end{equation}
\begin{equation}
    \dot{f_{a}}=\frac{C_{a}}{r_{a}^{2}}+\frac{\Lambda_{1}}{2}
\end{equation}Substituting the expression of $\dot{f_{a}}$ in the equation(\ref{Pulsating Lagrangian}) we get,
\begin{eqnarray}
\mathcal{L}=-\sqrt{2\lambda}\left[\frac{\dot{z_{0}}^{2}+z_{0}^{2}\dot{h_{0}}^{2}}{4}\right]+\sqrt{2\lambda}\sum_{a=1}^{4}\left[\dot{r_{a}}^{2}+\frac{C_{a}^{2}}{r_{a}^{2}}+C_{a}\Lambda_{1}+\frac{\Lambda_{1}^{2}}{4}r_{a}^{2}-r_{a}^{2}m_{a}^{2}\right]\nonumber\\+b\sqrt{2\lambda}\sum_{a=1}^{4}r_{a}\dot{r_{a}}m_{a}-\frac{\Lambda}{2}\sqrt{8\lambda}\left(\sum_{a=1}^{4}r_{a}^{2}-1\right)-\frac{\Tilde{\Lambda}}{2}\sqrt{8\lambda}\left(z_{0}^{2}+1 \right)\nonumber\\
-\sqrt{8\lambda}\frac{\Lambda_{1}}{2}\sum_{a=1}^{4}\left(C_{a}+\frac{\Lambda_{1}}{2}r_{a}^{2}\right)-\sqrt{8\Lambda}\frac{\Lambda_{2}}{2}\sum_{a=1}^{4}r_{a}^{2}m_{a}.
\end{eqnarray}
This eventually yields the equation of motion for $r_{a}$ as:
\begin{equation}
    \Ddot{r}_{a}+\frac{C_{a}^{2}}{r_{a}^{3}}+\frac{\Lambda_{1}^{2}}{4}+\left(\Lambda+\Lambda_{2}m_{a}+m_{a}^{2}\right)r_{a}^{2}=0.
    \label{r equation}
\end{equation}One can show that the equations of motion (\ref{z equation}) and (\ref{r equation}) can also be obtained from a Lagrangian 
\begin{eqnarray}
    L_{NR}=\frac{\dot{z_{0}}^{2}}{4}-\frac{C_{0}^{2}}{4z_{0}^{0}}-\Tilde{\Lambda}\left(z_{0}^{2}+1\right)+\sum_{a=1}^{4}\left(\dot{r}_{a}^{2}+\frac{C_{a}^{2}}{r_{a}^{2}}\right)+B\sum_{a=1}^{4}r_{a}\dot{r}_{a}m_{a}\nonumber\\+\sum_{a=1}^{4}\left(r_{a}^{2}-1\right)\left(\sum_{a=1}^{4}C_{a}\right)^{2} -\sum_{a=1}^{4}m_{a}^{2}r_{a}^{2}+\Lambda\left(\sum_{a=1}^{4}r_{a}^{2}-1\right)+\Lambda_{2}\sum_{a=1}^{4}m_{a}r_{a}^{2}.
    \end{eqnarray} It is obvious that this Lagrangian is of the form of a NR system with $z_{0}^{2}$ , $\frac{1}{z_{0}^{2}}$, $r_{a}^{2}$ and $\frac{1}{r_{a}^{2}}$ type of terms. Here we have used $\Lambda_{1}=\left(-2\sum_{a=1}^{4}C_{a}\right)$. 
    The Hamiltonian formulation yields 
    \begin{eqnarray}
        H_{NR}=\frac{\dot{z}_{0}^{2}}{4}+\frac{C_{0}^{2}}{4z_{0}^{2}}+\sum_{a=1}^{4}\left(\dot{r}_{a}^{2}-\frac{C_{a}^{2}}{r_{a}^{2}}\right)+\sum_{a=1}^{4}m_{a}^{2}r_{a}^{2}.
     \end{eqnarray} 
         The corresponding Virasoro constraints may be written as:
         \begin{equation}
         \sum_{a=1}^{4}\left[\dot{r}_{a}^{2}+r_{a}^{2}\dot{f}_{a}^{2}+r_{a}^{2}m_{a}^{2}\right]=\left(\frac{\dot{z}_{0}^{2}}{4}+\frac{C_{0}^{2}}{4z_{0}^{2}}\right),
         \label{virasoro 1}
         \end{equation}
         \begin{equation}
         \sum_{a=1}^{4}r_{a}^{2}\dot{f}_{a}m_{a}=0.
             \end{equation}
The Uhlenback integrals of motion involved in the motion of the pulsating string in $R_t \times \mathbb{CP}^{3}$ in the presence of flux may be obtained by using the similar procedure as described in section (2.3) and those are \footnote{here also we consider $b=1$ for simplicity.}
             \begin{eqnarray}
                 F=z_{0}^{2}+r_{a}^{2}\sum_{b\neq a}\frac{\left(\dot{r}_{a}r_{b}-r_{a}\dot{r}_{b}\right)^{2}}{m_{a}^{2}-m_{b}^{2}}+\sum_{b\neq a} \frac{1}{m_{a}^{2}-m_{b}^{2}}\left(\frac{C_{a}r_{b}}{r_{a}}+\frac{C_{b}r_{a}}{r_{b}}\right)^{2}\nonumber\\+2\sum_{b\neq a}\frac{\left(r_{a}\dot{r}_{a}r_{b}^{2}-r_{b}\dot{r}_{b}r_{a}^{2}\right)}{m_{a}+m_{b}}+\frac{1}{4}\sum_{b\neq a}\left(\frac{m_{a}-m_{b}}{m_{a}+m_{b}}\right)r_{a}^{2}r_{b}^{2}.
             \end{eqnarray}
             \subsection{String profile and conserved charges}
             To find the expression of $r_{a}(\tau)$, firstly we take $r_{1}$=$r_{3}$=$\frac{1}{\sqrt{2}}\sin{\theta}$, $r_{2}$=$r_{4}$=$\frac{1}{\sqrt{2}}\cos{\theta}$, $m_{1}=-m_{3}$ and $m_{2}=-m_{4}$, so that the constraints become
             \begin{equation}
             \sum_{a=1}^{4}r_{a}^{2}=1, \sum_{a=1}^{4}r_{a}^{2}m_{a}=0.
                \end{equation}Now if we consider $z_{0}(\tau)=1$ and $h_{0}(\tau)=\tau=t$, from the Virasoro constraint(\ref{virasoro 1}) we get 
                \begin{equation}
                    \dot{\theta}^{2}=\frac{1}{4}-\frac{2C_{1}^{2}}{\sin^{2}\theta}-\left(m_{1}^{2}sin^{2}\theta+m_{2}^{2}\cos^{2}\theta\right), 
                    \label{pulsating eom}
                    \end{equation}
                    with $C_{2}=C_{4}$=0 and $C_{1}=-C_{3}$.
                    Equation of motion(\ref{pulsating eom}) finally yields 
                    \begin{equation}
                        \frac{\partial}{\partial\tau}(\cos\theta)=\dot{r}_{2}=\mp \sqrt{m_{1}^{2}-m_{2}^{2}}\sqrt{\left(\cos^{2}\theta-z_{-}^{2}\right)\left(z_{+}^{2}-\cos^{2}\theta\right)}
                    \end{equation}
             which gives a solution for all $r_{a}$'s, $a=1,2,3,4$ as
             \begin{equation}
                 r_{1}=r_{3}=\frac{1}{\sqrt{2}}\sqrt{1-z_{+}^{2}dn^{2}\left(A\tau|m\right)};~~
                  r_{2}=r_{4}=\frac{z_{+}}{\sqrt{2}}dn\left(A\tau|m\right) ,
             \end{equation}where 
             \begin{equation}
                 A\equiv \mp z_{+}\tau\sqrt{m_{1}^{2}-m_{2}^{2}} ,  m \equiv 1-\frac{z_{-}^{2}}{z_{+}^{2}},
             \end{equation} 
             \begin{equation}
             z_{\pm}^{2}=\frac{1}{2\left(m_{1}^{2}-m_{2}^{2}\right)}\left[\left(\frac{1}{4}-2m_{1}^{2}+m_{2}^{2}\right)\pm \sqrt {\left(\frac{1}{4}-2m_{1}^{2}+m_{2}^{2}\right)^{2}-4\left(m_{1}^{2}-m_{2}^{2}\right)\left(m_{1}^{2}+2C_{1}^{2}-\frac{1}{4}\right)}\right].
                \end{equation}Hence the string solutions may be written as
             \begin{subequations}
             \begin{align}
             Z&=\frac{R}{2}e^{i\tau},~~
              W_{1}=\frac{R}{\sqrt{2}}\sqrt{1-z_{+}^{2}dn^{2}(A\tau|m)}e^{i(m_{1}\sigma+f_{1})},\\
 W_{2}&=\frac{R}{\sqrt{2}}z_{+}dn(A\tau|m)e^{i(m_{2}\sigma+f_{2})},~~
W_{3}=\frac{R}{\sqrt{2}}\sqrt{1-z_{+}^{2}dn^{2}(A\tau|m)}e^{-i(m_{1}\sigma+f_{1})},\\
 W_{4}&=\frac{R}{\sqrt{2}}z_{+}dn(A\tau|m)e^{-i(m_{2}\sigma+f_{2})}.
                 \end{align}
             \end{subequations}Now, the conserved charges for the pulsating string in such a background may be found from the target space Lagrangian as
             \begin{equation}
             E_{s}=\frac{\partial\mathcal{L}}{\partial(\partial_{\tau}t)}=-\frac{\sqrt{2\lambda}}{2},~~J_{a}=\frac{\partial\mathcal{L}}{\partial(\partial_{\tau}\phi_{a})}=-\sqrt{2\lambda}r_{a}^{2}\dot{f}_{a}
            \end{equation}
            It is obvious that the presence of flux in the background does not affect the conserved charges in the case of a pulsating string. Now expressing the Virasoro constraint(\ref{virasoro 1}) in terms of $\mathcal{E}=\frac{E}{\sqrt{2\lambda}}$ and $\mathcal{J}_{a}=\frac{J_{a}}{\sqrt{2\lambda}}$, the oscillation number can be written as \cite{Pradhan:2013sja}
            \begin{equation}
                \mathcal{N}=\frac{N}{\sqrt{2\lambda}}=\oint r_{1}\dot{r}_{1}dr_{1}=\sqrt{2\lambda}\int_{0}^{\sqrt{R}}dr_{1}\frac{1}{2r_{1}}\sqrt{4r_{1}^{2}\mathcal{E}^{2}-\mathcal{J}_{1}^{2}-4m_{1}^{2}r_{1}^{4}}.
            \end{equation}where $\sqrt{R}=r_{1max}=\frac{1}{m_{1}}\left[\mathcal{E}^{2}\pm \sqrt{\mathcal{E}^{2}-\mathcal{J}_{1}^{2}m_{1}^{2}}\right]^{\frac{1}{2}}$.
            From the above we get,
            \begin{equation}
                \frac{\partial\mathcal{N}}{\partial m_{1}}=-2m_{1}\int_{0}^{\sqrt{R}}\frac{r_{1}^{3}}{\sqrt{\left(r_{1}^{2}-\frac{a_{-}}{m_{1}^{2}}\right)\left(\frac{a_{+}}{m_{1}^{2}}-r_{1}^{2}\right)}}dr_{1},
            \end{equation}where $a_{\pm}=\mathcal{E}^{2}\pm\sqrt{\mathcal{E}^{2}-\mathcal{J}_{1}^{2}m_{1}^{2}}$ and $\sqrt{R}=\sqrt{\frac{a_{+}}{m_{1}^{2}}}$. Expressing it in terms of standard elliptic integrals and dropping some terms for sake of simplicity we get,
            \begin{equation}
              \frac{\partial\mathcal{N}}{\partial m_{1}}=2\sqrt{m_{1}}a_{+}^{\frac{3}{2}}\left[\textbf{K}\left(\frac{a_{-}}{a_{+}}\right)-\textbf{E}\left(\frac{a_{-}}{a_{+}}\right)\right].
            \end{equation}
            Now we use the following standard expansions of the elliptic integrals of first and second kinds, 
            \begin{equation}
                \textbf{K}\left(\epsilon\right)=\frac{\pi}{2}+\frac{\pi\epsilon}{8}+\frac{9\pi\epsilon^{2}}{128}+\frac{25\pi\epsilon^{3}}{512}+\frac{1225\pi\epsilon^{4}}{32768}+ \mathcal{O}[\epsilon]^{5},
            \end{equation}
            \begin{equation}
                \textbf{E}\left(\epsilon\right)=\frac{\pi}{2}-\frac{\pi\epsilon}{8}-\frac{3\pi\epsilon^{2}}{128}-\frac{5\pi\epsilon^{3}}{512}-\frac{175\pi\epsilon^{4}}{32768}+\mathcal{O}[\epsilon]^{5},
            \end{equation}where $\epsilon=\frac{a_{-}}{a_{+}}$.
            Substituting the expressions for $a_{+}$ and $a_{-}$ in the above expansion and expanding for  small $\mathcal{E}$ and $\mathcal{J}_{1}$, we get
            \begin{equation}
            \begin{split}
            \frac{\partial\mathcal{N}}{\partial m_{a}}=&\left(\frac{11}{64}\pi\mathcal{J}_{a}m_{a}^{\frac{5}{2}}-\frac{135}{1024}\pi\mathcal{J}_{a}^{4}m_{a}^{\frac{9}{2}}+\frac{175}{2048}\pi\mathcal{J}_{a}^{6}m_{a}^{\frac{13}{2}}+\mathcal{O}[\mathcal{J}_{a}^{8}]\right)\\& -\left(\frac{\pi}{8}m_{a}^{\frac{1}{2}}+\frac{117\pi}{512}\mathcal{J}_{a}^{2}m_{a}^{\frac{5}{2}}-\frac{175}{4096}\pi\mathcal{J}_{a}^{6}m_{a}^{\frac{9}{2}}+\mathcal{O}[\mathcal{J}_{a}^{8}]\right)\mathcal{E}
            \\&
            +\left(-\frac{3}{32}\pi m_{a}^{\frac{1}{2}}+\frac{21}{128}\pi\mathcal{J}_{a}^{2}m_{a}^{\frac{5}{2}}-\frac{525}{8192}\pi\mathcal{J}_{a}^{4}m_{a}^{\frac{9}{2}}+\mathcal{O}[\mathcal{J}_{a}^{6}]\right)\mathcal{E}^{2}\\& -\left(\frac{3}{32}m_{1}^{\frac{1}{2}}-\frac{57}{128}\pi\mathcal{J}_{a}^{2}m_{a}^{\frac{5}{2}}+\frac{525}{2048}\pi\mathcal{J}_{a}^{4}m_{a}^{\frac{9}{2}}+\mathcal{O}[\mathcal{J}_{a}^{6}]\mathcal{E}^{3}\right)+\mathcal{O}[\mathcal{E}^{4}].
            \end{split}
            \label{OM1}
                \end{equation}
            Integrating the equation(\ref{OM1}) with respect to $m$ and then inverting the series we get the energy as a function of oscillation number, winding number and conserved angular momenta as,
            \begin{equation}
                \mathcal{E}=\mathcal{M}+K(m_{a})\frac{5\pi m_{a}}{32}\mathcal{J}_{a}^{2}+\mathcal{O}[\mathcal{J}_{a}]^{4},
            \end{equation}where,
            \begin{eqnarray}
                \mathcal{M}=&\frac{\pi}{16\sqrt{m_{a}}}+\sqrt{\mathcal{N}},\nonumber\\
                K(m_{a})=&\frac{117m_{a}^{\frac{1}{2}}}{32}-\mathcal{N}\left(11m_{a}^{\frac{1}{2}}+\frac{63\pi}{256}\right).
            \end{eqnarray}When we take $m_{a}\rightarrow 1$ and $\mathcal{J}_{a}\rightarrow 0$ it yields the first order term of the small energy limit expansion of the energy for the string pulsating in one plane\cite{Park:2005kt,Cardona:2014gqa}.
        \section{Conclusions and Outlook}
We have shown, in this paper, that the string motion in $AdS_4 \times {\mathbb{CP}}^3$ in the presence of $B_{NS}$ holonomy is integrable by reducing the Lagrangian of such a system into the form of a NR model with an additional term proportional to the flux. We have elucidated the Lagrangian and Hamiltonian formulation and computed the integrals of motion in our pursuit to show it to be integrable. We have then turned out attention for studying the rotating string with two angular momenta and have found out the relevant scaling relation among various charges corresponding to the giant magnon solution of string rotating in this background. We have also computed the leading order finite size correction of such dispersion relation. For the pulsating string we have performed the Lagrangian and Hamiltonian formulation and integrals of motion to show that it reduces to a NR system. We have derived the integrable equations of motion, the pulsating string profile and the short string energy as function of oscillation number and angular momenta. It would be certainly interesting to generalize the construction to the more generic rotating and pulsating string solutions and check the integrability. The presence of $B_{NS}$ field does not destroy the integrability of the background and hence it will be interesting to check for general rotating strings in the ABJ background. It would also be interesting to look at the D1-string equation of motion in this background and check the integrability via the NR system. We wish to come back to some of these issues in future. It will also be interesting to look for finite-size corrections by using the L\"{u}scher correction formulation based on exact S-matrix.

\end{document}